\documentclass[12pt]{article}
\usepackage{amssymb, amsmath, bm, float, epsfig, wrapfig, color}

\begin{document}
\title{
\begin{flushright}
\ \\*[-80pt]
\begin{minipage}{0.2\linewidth}
\normalsize
KIAS-P16014
HUPD1605
WU-HEP-16-04 \\*[50pt]
\end{minipage}
\end{flushright}
{\Large \bf 
Double Froggatt{--}Nielsen Mechanism
\\*[20pt]}
}

\author{
Kenji~Nishiwaki,$^1$\footnote{E-mail address: nishiken@kias.re.kr} \quad
Yusuke~Shimizu,$^{1,2,3}$\footnote{E-mail address: yu-shimizu@hiroshima-u.ac.jp} \, and \,
Yoshiyuki~Tatsuta$^4$\footnote{E-mail address: y\_tatsuta@akane.waseda.jp}\\*[30pt]
$^1${\it \normalsize School~of~Physics,~KIAS,~Seoul~130-722,~Republic~of~Korea}\\
$^2${\it \normalsize Quantum~Universe~Center,~KIAS,~Seoul~130-722,~Republic~of~Korea}\\
$^3${\it \normalsize Graduate~School~of~Science,~Hiroshima~University,~Higashi-Hiroshima,~739-8526,~Japan}\\
$^4${\it \normalsize Department~of~Physics,~Waseda~University,~Tokyo~169-8555,~Japan}\\*[55pt]
}

\date{
\centerline{\small \bf Abstract}
\begin{minipage}{0.9\textwidth}
\medskip 
\medskip 
\small 
We present a doubly parametric extension of the standard Froggatt{--}Nielsen (FN) mechanism. 
{As is well known,} mass matrices of the up- and down-type quark sectors and the charged lepton sector 
in the standard model can be parametrized {well} by a parameter $\lambda$ which is usually taken to be the sine of the Cabibbo angle 
($\lambda = {\sin{\theta_{\text{C}}}} \simeq 0.225$). 
However, in the neutrino sector, there is still room to realize the two neutrino mass squared differences 
$\Delta m_\text{sol}^2$ and $\Delta m_\text{atm}^2$, two large mixing angles $\theta _{12}$ and $\theta _{23}$, and
non-zero $\theta _{13}$.
Then we consider an extension with an additional parameter $\rho$ in addition to $\lambda$. 
Taking the relevant FN charges for {a} power of $\lambda~(=0.225)$ and additional FN charges for {a} power of $\rho$, which we assume {to be} less than one, 
we can reproduce the ratio of the two neutrino mass squared differences and three mixing angles.
In the normal neutrino mass hierarchy, we show several patterns for taking relevant FN charges and {the} magnitude of $\rho$. 
We find that if $\sin \theta_{23}$ is measured more precisely, we can distinguish each pattern. 
{This} is testable in the near future, {for example} in neutrino oscillation experiments. 
In addition, we predict the Dirac {CP-violating} phase for each pattern. 
\end{minipage}}

\begin{titlepage}
\maketitle
\thispagestyle{empty}
\end{titlepage}


\section{Introduction}
The standard model (SM) is one of the successful models {in explaining} the results of recent precise experiments. 
However, there are many free parameters{, particularly} as Yukawa couplings in the SM. There are some ambiguities {in realizing} 
the quark and lepton mass hierarchies and mixing angles. 
Then{,} many authors have studied texture analyses or flavor symmetry models 
in order to elucidate the origin of the flavor structure as a direction beyond the SM. 
{In fact}, Weinberg proposed a simple {zero} texture, within two generations of quarks, 
where quark masses and a mixing angle are related~\cite{Weinberg:1977hb}. 
Fritzsch extended {this} to three generations {in the} so-called ``Fritzsch-type mass matrix"~\cite{Fritzsch:1977vd,Fritzsch:1979zq}{,}
which relates quark masses and mixing angles in the quark sector. 
Furthermore, Fukugita, Tanimoto, and Yanagida extended this argument to the lepton sector~\cite{Fukugita:1992sy} and predicted two large neutrino mixing angles 
and non-zero $\theta _{13}$~\cite{Fukugita:2003tn,Fukugita:2012jr}{,} which was the last mixing angle of the lepton sector~\cite{An:2012eh}-\cite{Abe:2014lus}.  

On the other hand, flavor symmetry also plays an important role {in understanding} the flavor structure. 
Froggatt and Nielsen proposed {the} so-called ``Froggatt{--}Nielsen (FN) mechanism"~\cite{Froggatt:1978nt}{,} which introduces {$U(1)_{\text{FN}}$} symmetry as flavor symmetry. 
Taking relevant {$U(1)_{\text{FN}}$} charge assignments to the different generations, 
the quark mass hierarchy and {Cabibbo--Kobayashi--Maskawa} (CKM) matrix are naturally reproduced in the quark sector, 
while the non-Abelian discrete flavor symmetries~\cite{Ishimori:2010au}-\cite{King:2014nza} can easily derive the large mixing angles in the lepton sector, 
e.g., tri-bimaximal (TBM) mixing~\cite{Harrison:2002er,Harrison:2002kp}, which is a mixing paradigm by describing simple mass textures. 
After the reactor neutrino experiments reported non-zero $\theta _{13}$, it is important to study other flavor paradigms, 
e.g., tri-bimaximal{--}Cabibbo (TBC) mixing~\cite{King:2012vj,Shimizu:2012ry}{,} with the same mindset as TBM, bi-maximal (BM), tri-maximal, 
and golden ratio neutrino mixing.
{(For a review, see~\cite{Ishimori:2010au}-\cite{King:2014nza})}. 
Thus the texture analysis and the flavor symmetry are important {in understanding} the flavor structures for {both} quark and lepton sectors 
and there are many works known as, e.g., ``stitching the Yukawa quilt"~\cite{Ramond:1993kv}, ``$\mu$-$\tau $ anarchy"~\cite{Fukuyama:1997ky,Buchmuller:1998zf}, 
``cascades"~\cite{Haba:2008dp}, ``Occam's razor"~\cite{Harigaya:2012bw,Tanimoto:2016rqy}, and ``repressing anarchy"~\cite{Altarelli:2012ia,Bergstrom:2014owa}. 

{As is well known}, mass matrices of the up- and down-type quarks and the charged leptons in the SM 
can be {parametrized well} by a parameter $\lambda${,} which is usually taken to be the sine of the Cabibbo angle. 
Taking $\lambda \simeq 0.225$, up- and down-type quark and charged lepton mass hierarchies and mixing angles are reproduced. 
This type of parametrization was originally proposed by Froggatt and Nielsen~\cite{Froggatt:1978nt}. 
On the other hand, however, in the neutrino sector, there is still room to realize the lepton flavor structure, i.e., 
neutrino mass squared differences $\Delta m_\text{sol}^2$ and $\Delta m_\text{atm}^2$, 
two large mixing angles $\theta _{12}$ and $\theta _{23}$, and non-zero $\theta _{13}$.
Indeed, it is likely that the neutrino mass matrix has a {property distinct} from those of the {up- and down-type quarks} and the charged lepton mass matrices.
This is due to the fact that the neutrino masses are so tiny in comparison with the other SM fermion masses, 
and that the lepton mixing angles are relatively larger than the quark mixing angles. 
In this paper, we present an extension of the FN mechanism in the neutrino mass matrix.
In particular, we focus on a doubly parametric extension of the FN (DFN) mechanism.\footnote{In some models of the up- and down-type quark sectors, 
it is known that the mass matrices are consequently parametrized by two parameters. For example, see Refs.~\cite{Leurer:1992wg, Leurer:1993gy}.} 
{To explain}, we show an illustrative example of the doubly parametric extension.
This extension is plausible when we use the seesaw mechanism~\cite{Minkowski:1977sc}-\cite{Schechter:1981cv}, for instance.
If the neutrinos are Majorana particles, in the seesaw mechanism, 
we need {both} Dirac and Majorana mass terms. Even if the Dirac neutrino mass matrix is parametrized by $\lambda$ like the other SM fermions, 
where the Dirac-type masses come from spontaneous symmetry breaking in the SM, 
the Majorana masses can include free mass parameters in general, and in some models it is plausible that Majorana masses are parametrized by another parameter. 
Then in the neutrino sector, such a situation corresponds to an extended FN mechanism with {a parameter} $\rho$ 
in addition to $\lambda$.\footnote{Note that we show another extension of {the} FN mechanism{, the} Gaussian FN mechanism on magnetized orbifolds in Ref.~\cite{Abe:2014vza}.} 
Taking the relevant FN charges for {the} power of $\lambda~(=0.225)$ and additional FN charges for {the} power of $\rho${,} which we assume {to be} less than one, 
we can reproduce the ratio of two neutrino mass squared differences and three mixing angles. 
In our numerical calculations, we show several patterns for taking relevant FN charges and {the magnitude of} $\rho$. 
Note that in our numerical analyses, we consider only {the} normal neutrino mass hierarchy. 
We find that if $\sin \theta _{23}$ is measured more precisely, we can distinguish each pattern. 
In addition, we predict the Dirac {CP-violating} phase ({$\delta _{\text{CP}}$}) for each pattern. 

This paper is organized as follows. 
In {Sect.}~\ref{sec:DFN}, we show the standard FN mechanism and present the DFN mechanism. 
In {Sect.}~\ref{sec:analyses}, we show the results of our numerical analyses in several patterns. 
{Section}~\ref{sec:summary} is devoted to discussions and summary. 
In {Appendix}~\ref{sec:mass_matrices}, we show the explicit form of the neutrino mass matrix for each pattern.


\section{Doubly parametric extension of the FN mechanism}
\label{sec:DFN}
It is known that {the} mass matrices of the up- and down-type quark sectors and the charged lepton sector in the SM 
can be {parametrized well} by a parameter $\lambda$ and six charges $\{a_i, b_j\}~(i,j=1,2,3)$, i.e.,
\begin{equation}
m_{ij} = \lambda ^{a_i + b_j},
\label{eq:SFN}
\end{equation}
{with} up to ${\cal O}(1)$ complex coefficients in front of each element.\footnote{
We note that this form can be derived from extra dimensions.
When we assume that the SM fermions propagate in the bulk of an interval and have Yukawa couplings on the brane at $y=L$, 
the mass matrix is symbolically written down as $m_{ij} \propto e^{L (M_{L_{i}} - M_{R_{j}})}\sim \lambda^{a_{i} + b_{j}}$, where $L$ {is} the length of the interval. 
$M_{L_{i}/R_{i}}$ are bulk masses for $i$th-generation doublet/singlet, 
and we put the Dirichlet boundary condition for right/left-handed mode of the doublets/singlets at the two end points at $y=0,L$ 
{to realize} left-hand doublet modes and right-hand singlet modes, respectively. 
We adopt the notation in Ref.~\cite{Fujimoto:2016gfu}. 
If the particle profiles are also localized among other directions of extra dimensions, we might address the DFN structure.} 
In particular, it is reasonable to choose a value of the parameter $\lambda$ such that {the} observed masses and mixing angles of 
the up- and down-type quark and the charged lepton sectors can be realized. 
Indeed, such a value is given as $\lambda = {\sin \theta_{\text{C}}} \simeq 0.225$, where {$\theta_{\text{C}}$} is the Cabibbo angle.
This type of parametrization was originally proposed by Froggatt and Nielsen, {the} so-called ``FN parametrization"~\cite{Froggatt:1978nt}. 
 
In this paper, we consider an extension with an additional parameter $\rho$ and six additional charges $\{c_i, d_j\} ~ (i,j=1,2,3)$, i.e.,
\begin{equation}
m_{ij} = \lambda^{a_i + b_j} \rho^{c_i+d_j},
\label{eq:DFN}
\end{equation}
{also with} up to ${\cal O}(1)$ complex coefficients in front of each element.
In particular, this parametrization is valid in a neutrino mass matrix as well as the other fermion mass matrices. 
It should be noted that there are some possibilities {for realizing} the DFN parametrization.
For example, the DFN parametrization would be considered as effective theories of multi-scale extra dimensions, and would be obtained by an additional $U(1)$ flavor symmetry and so on.
To construct concrete models including the DFN parametrization is beyond the scope of this paper.
In {Refs}.~\cite{Leurer:1992wg, Leurer:1993gy}, for the up- and down-type quark sectors, the phenomenological prospects of the doubly parametric extension have {already been} studied.
In this paper, we focus only on phenomenological properties of the doubly parametric extension in the lepton sector, {in particular} the neutrino mass matrix. 
Here, we assume that the charged lepton mass matrix takes a diagonal form. 

{Finally}, it is important to comment on concrete values of {the} two parameters $\lambda$ and $\rho$.
Note that without loss of generality we can choose the value of {the} original parameter $\lambda$ such that $\lambda=0.225$.
Even if the parameter is chosen to be {a} distinct value, we can move to the case of $\lambda=0.225$ by redefining {the} additional FN charges $\{c_i, d_j\}$ and the value of {the} additional parameter $\rho$.
Hence, in the following, we take $\lambda =0.225$ and $\rho$ as an arbitrary value which we assume {to be} less than one. 
We show that the DFN textures can reproduce the ratio of the two neutrino mass squared differences and three mixing angles. 
We also show the results of our numerical analyses in the next section. 
Here, we do not identify the origin of the additional parameter $\rho$, where {one} possibility is (right-handed) Majorana neutrino mass parameters in a seesaw model.


\section{Numerical analyses}
\label{sec:analyses}
In this section, we focus on mass matrices in the neutrino sector, and also analyze numerical aspects of the extended FN parametrization. 
In our numerical calculations, we assume {the} normal neutrino mass hierarchy. 
We use the {results} of the global analysis of neutrino oscillation experiments~\cite{Gonzalez-Garcia:2015qrr}. 
The {$3\,\sigma$} ranges of the experimental data for the normal neutrino mass hierarchy are given as
\begin{align}
0.270\leq \sin ^2\theta _{12}&\leq 0.344,\quad 0.382\leq \sin ^2\theta _{23}\leq 0.643,\quad 0.0186\leq \sin ^2\theta _{13}\leq 0.0250, \nonumber \\
7.02&\leq \frac{\Delta m_\text{sol}^2}{10^{-5}~\text{eV}^2}\leq 8.09,\quad 2.317\leq \frac{\Delta m_\text{atm}^2}{10^{-3}~\text{eV}^2}\leq 2.607,
\label{eq:neutrino_data}
\end{align}
where {$\theta _{ij}$} are lepton mixing angles in the {Pontecorvo--Maki--Nakagawa--Sakata} (PMNS) matrix, 
while $\Delta m_\text{sol}^2$ and $\Delta m_\text{atm}^2$ are the solar and atmospheric neutrino mass squared differences, respectively. 
Note that the DFN parametrization can be valid in {both} Dirac and Majorana mass matrices.
For simplicity, we assume that FN charges $a_i (c_i)$ are equivalent to the other charges $b_j (d_j)$, respectively.
Now, the mass matrix in Eq.~(2) becomes symmetric.
We also assume that the mass matrix of the charged leptons is diagonal.\footnote{It is clear that we can consider large contributions to lepton mixing angles, e.g., Ref.~\cite{Antusch:2005kw}.
However, such large contributions are not typical setups in the framework of the FN parametrization.
In many typical cases, the contributions from the charged lepton mass matrices are considered to be small enough to be negligible.}

In the following, we consider 
six patterns\footnote{We analyzed various textures and 
{picked} six patterns to be mentioned where the result of the recent neutrino oscillation experiments~\cite{Gonzalez-Garcia:2015qrr} 
tends to be explained in part of parameter space.} 
with/without additional FN charges 
as sample patterns {for} numerical calculation{:}\footnote{We show the explicit form of the mass matrix of neutrinos for each pattern in {Appendix}~\ref{sec:mass_matrices}.} 
\begin{itemize}
\item Pattern 1 {:} $a_i=\{1,0,0\}${,} $\forall c_i$, and $\rho =1.0${;}
\item Pattern 2 {:} $a_i=\{1,0,0\}$, $c_i=\{0,\frac{3}{2},\frac{5}{2}\}$, and $\rho =0.8${;}
\item Pattern 3 {:} $a_i=\{\frac{3}{2},\frac{1}{2},0\}$, $c_i=\{0,\frac{1}{2},\frac{3}{2}\}$, and $\rho =0.4${;}
\item Pattern 4 {:} $a_i=\{\frac{3}{2},\frac{1}{2},0\}$, $c_i=\{0,\frac{1}{2},\frac{3}{2}\}$, and $\rho =0.5${;}
\item Pattern 5 {:} $a_i=\{\frac{3}{2},\frac{1}{2},0\}$, $c_i=\{0,\frac{1}{2},\frac{3}{2}\}$, and $\rho =0.6${;}
\item Pattern 6 {:} $a_i=\{\frac{1}{2},\frac{1}{2},0\}$, $c_i=\{1,0,\frac{1}{2}\}$, and $\rho =0.3$.
\end{itemize}
Note that with $\rho =1.0$, the first pattern of charge configurations gives the standard FN parametrization{,}
and this charge configuration gives a {$\mu$--$\tau$} symmetric mass matrix which derives the almost BM mixing. 
In our calculations, the three neutrino masses are adjusted by the ratio of the two neutrino mass squared differences, 
because an overall mass scale is completely free for our parametrization. Here we take ${\cal O}(1)$ coefficients 
as {$10\%$} deviations from unity and complex phases are taken from $-\pi$ to $\pi$.  
Then, we can predict the Dirac {CP-violating} phase {$\delta _{\text{CP}}$} in our numerical calculations, 
where the non-zero {$\delta _{\text{CP}}$} originates from the complex phases of the mass matrix elements. 
In each pattern, we scan $10^{6}$ configurations of the coefficients of the nine elements of the mass matrix.

\begin{figure}[H]
\begin{tabular}{c}
\begin{minipage}{0.32\hsize}
\centering
\includegraphics[clip, width=\hsize]{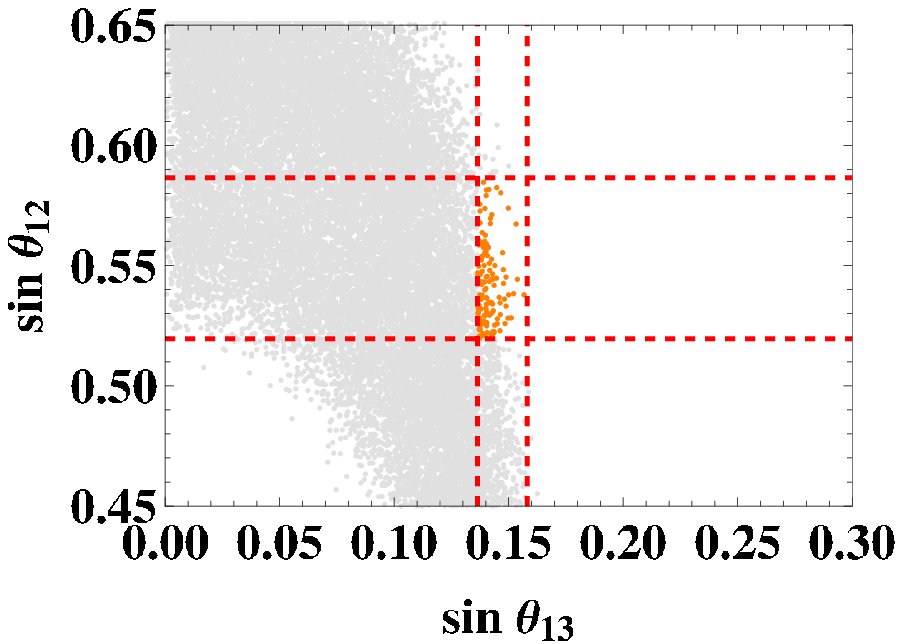}
\end{minipage}
\begin{minipage}{0.32\hsize}
\centering
\includegraphics[clip, width=\hsize]{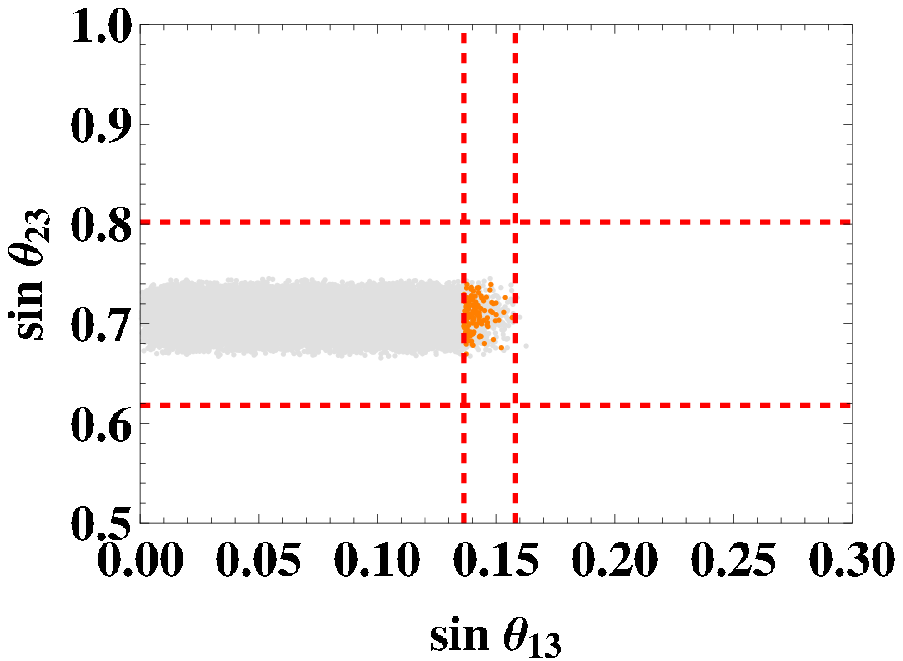}
\end{minipage}
\begin{minipage}{0.32\hsize}
\centering
\includegraphics[clip, width=\hsize]{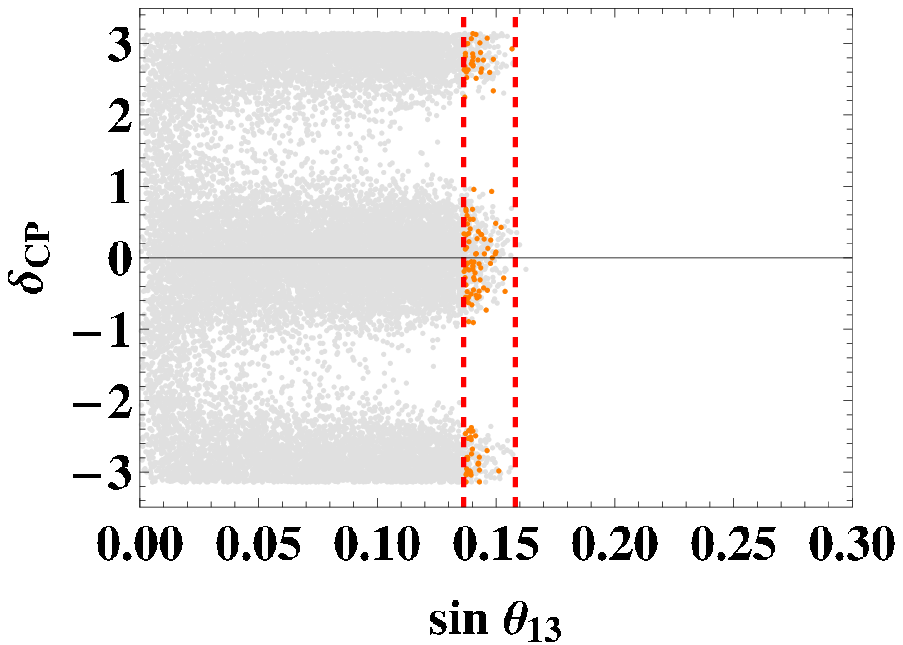}
\end{minipage}
\end{tabular}
\caption{Scatter plots in {Pattern} 1: $a_i=\{1,0,0\}${,} $\forall c_i$. We set $\rho =1.0$. 
The ratio of the two neutrino mass squared differences is required within the range where $\Delta m_{\rm sol}^2$ and $\Delta m_{\rm atm}^2$ are inside the {$3\,\sigma$} ranges shown in Eq.~(\ref{eq:neutrino_data}).
The insides of {the} red dotted lines show the {$3\,\sigma$} allowed regions of {the} lepton mixing angles. 
The orange points correspond to the case that {all} three lepton mixing angles 
are within the {$3\,\sigma$} ranges {simultaneously} in Eq.~(\ref{eq:neutrino_data}).}
\label{fig:pattern1}
\end{figure}

First, we show the scatter plots in {Pattern} 1. The gray regions suggest realized values of mixing angles $\sin \theta _{12}$, 
$\sin \theta _{23}$, $\sin \theta _{13}$, and Dirac {CP-violating} phase {$\delta _{\text{CP}}$} in Fig.~\ref{fig:pattern1}. 
The insides of {the} red dotted lines show the {$3\,\sigma$} allowed regions of each lepton mixing angle, while the orange points correspond to the case 
that {all} three lepton mixing angles are within the {$3\,\sigma$} ranges {simultaneously} in Eq.~(\ref{eq:neutrino_data}). 
We find that {$\delta _{\text{CP}}$} is predicted as $\left| {\delta _{\text{CP}}} \right| \lesssim 1$ and 
$2.2\lesssim \left| {\delta _{\text{CP}}} \right|\lesssim \pi$.

\begin{figure}[H]
\begin{tabular}{c}
\begin{minipage}{0.32\hsize}
\centering
\includegraphics[clip, width=\hsize]{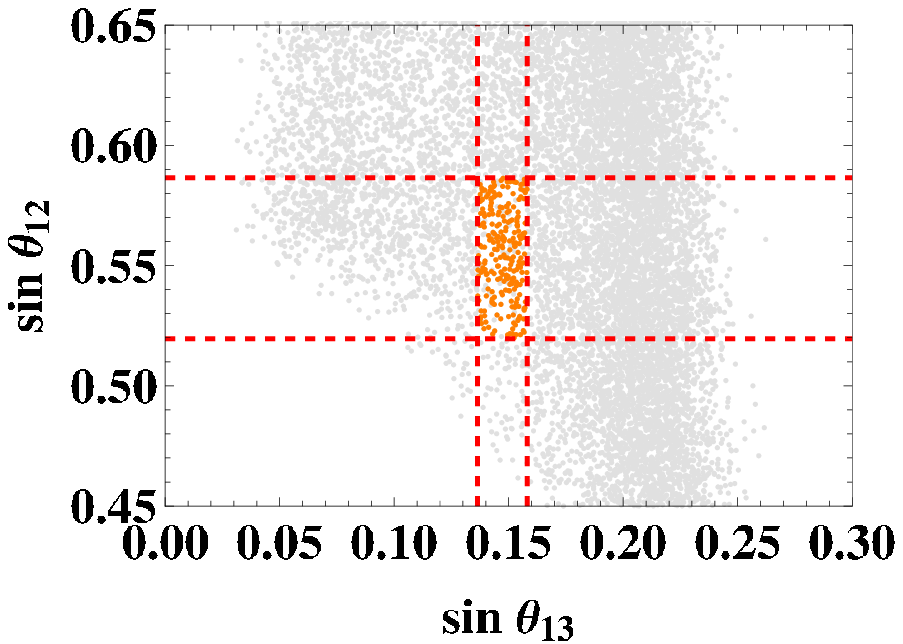}
\end{minipage}
\begin{minipage}{0.32\hsize}
\centering
\includegraphics[clip, width=\hsize]{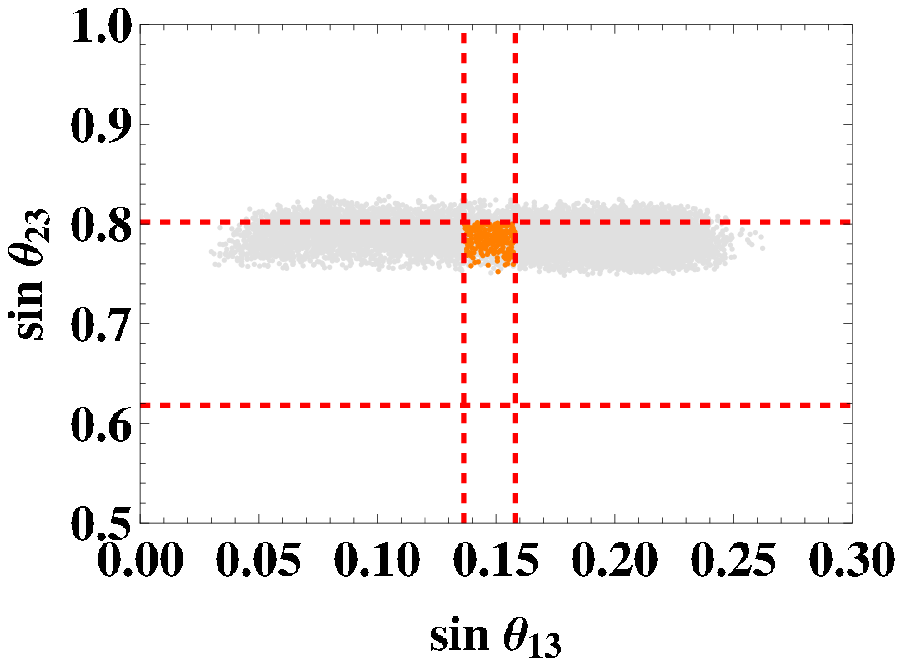}
\end{minipage}
\begin{minipage}{0.32\hsize}
\centering
\includegraphics[clip, width=\hsize]{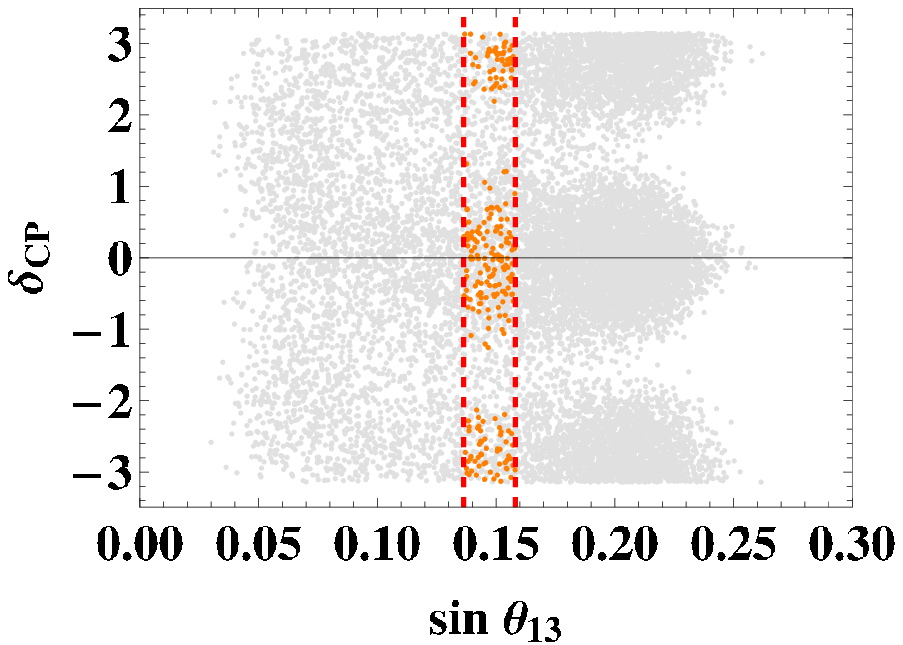}
\end{minipage}
\end{tabular}
\caption{Scatter plots in {Pattern} 2: $a_i=\{1,0,0\}$, $c_i=\{0,\frac{3}{2},\frac{5}{2}\}$. We set $\rho =0.8$. 
The color convention is {as in} Fig.~\ref{fig:pattern1}.}
\label{fig:pattern2}
\end{figure}

On the other hand, in {Pattern} 2, the standard FN charges are the same as those of {Pattern} 1 and we set $\rho =0.8$ so that the 
mass matrix becomes almost {$\mu$--$\tau$ symmetric (though not exactly).}
{Figure}~\ref{fig:pattern2} shows that $\sin \theta _{23}$ is around the upper boundary of the {$3\,\sigma$} range, 
while the other mixing angles are completely filled within the {$3\,\sigma$} range. 
{Comparing Figures~\ref{fig:pattern1} and \ref{fig:pattern2}}, it is easily {seen that the} realized values of $\sin \theta_{13}$ in {Pattern} 2 are relatively larger than those in {Pattern} 1.
This implies that $\sin \theta_{13}$ is improved by the DFN parametrizations, even if the coefficients are not so scattered in large parameter regions.
Also, we can distinguish {Patterns} 1 and 2 {between} the standard FN and DFN by more precise measurement of $\sin \theta _{23}$.
The extension with an additional parameter leads to modestly different properties of $\sin \theta_{13}$ and $\sin \theta_{23}$ from those of a {$\mu$--$\tau$} symmetric neutrino mass matrix.

\begin{figure}[H]
\begin{tabular}{c}
\begin{minipage}{0.32\hsize}
\centering
\includegraphics[clip, width=\hsize]{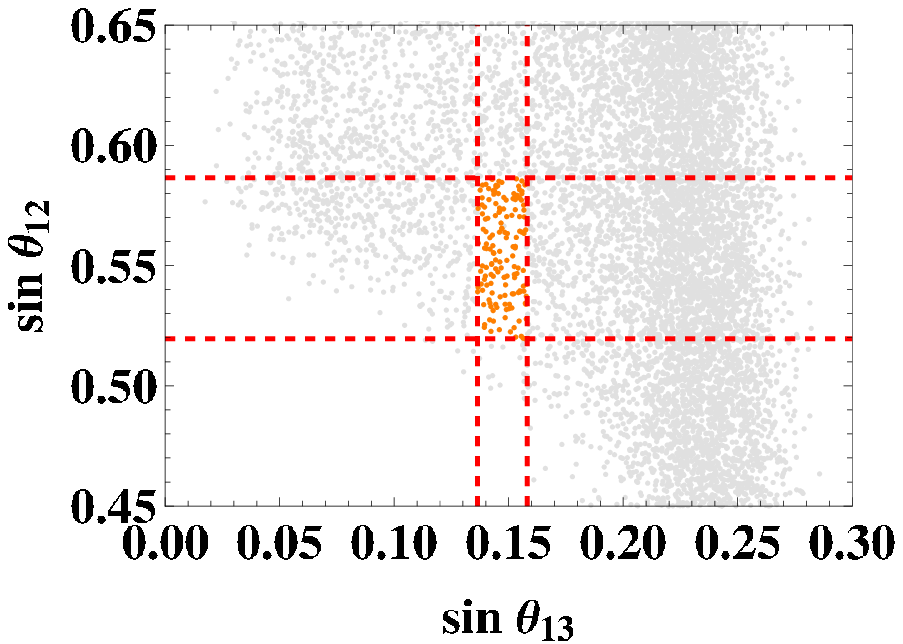}
\end{minipage}
\begin{minipage}{0.32\hsize}
\centering
\includegraphics[clip, width=\hsize]{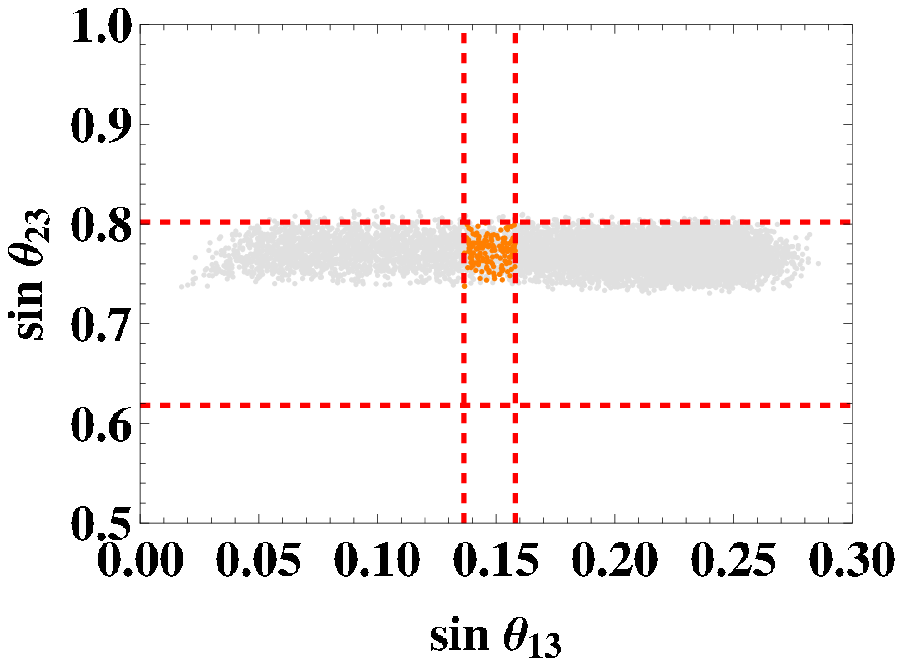}
\end{minipage}
\begin{minipage}{0.32\hsize}
\centering
\includegraphics[clip, width=\hsize]{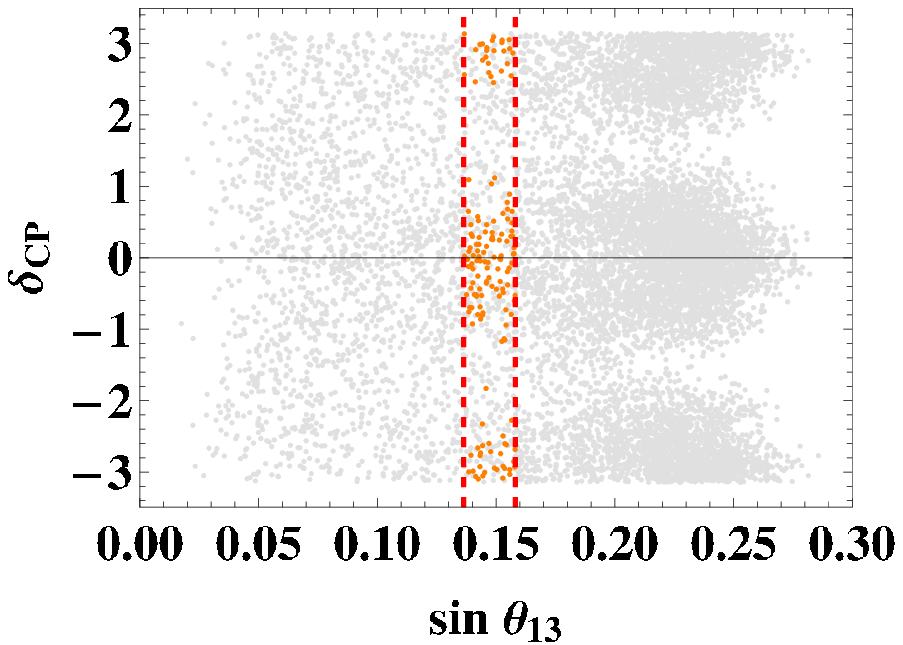}
\end{minipage}
\end{tabular}
\caption{Scatter plots in {Pattern} 3: $a_i=\{\frac{3}{2},\frac{1}{2},0\}$, $c_i=\{0,\frac{1}{2},\frac{3}{2}\}$. We set $\rho =0.4$. 
The color convention is {as in} Fig.~\ref{fig:pattern1}.}
\label{fig:pattern3}
\end{figure}

\begin{figure}[H]
\begin{tabular}{c}
\begin{minipage}{0.32\hsize}
\centering
\includegraphics[clip, width=\hsize]{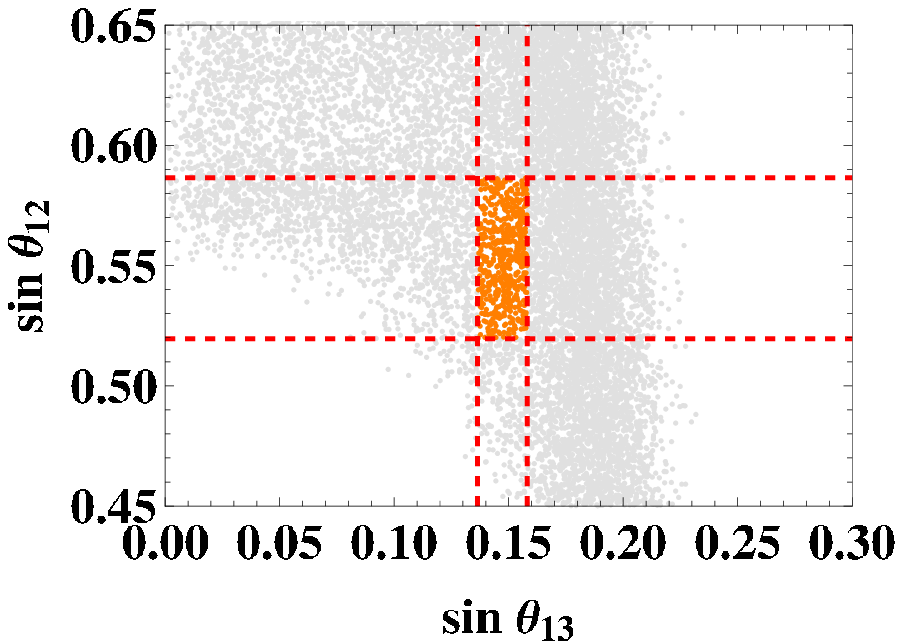}
\end{minipage}
\begin{minipage}{0.32\hsize}
\centering
\includegraphics[clip, width=\hsize]{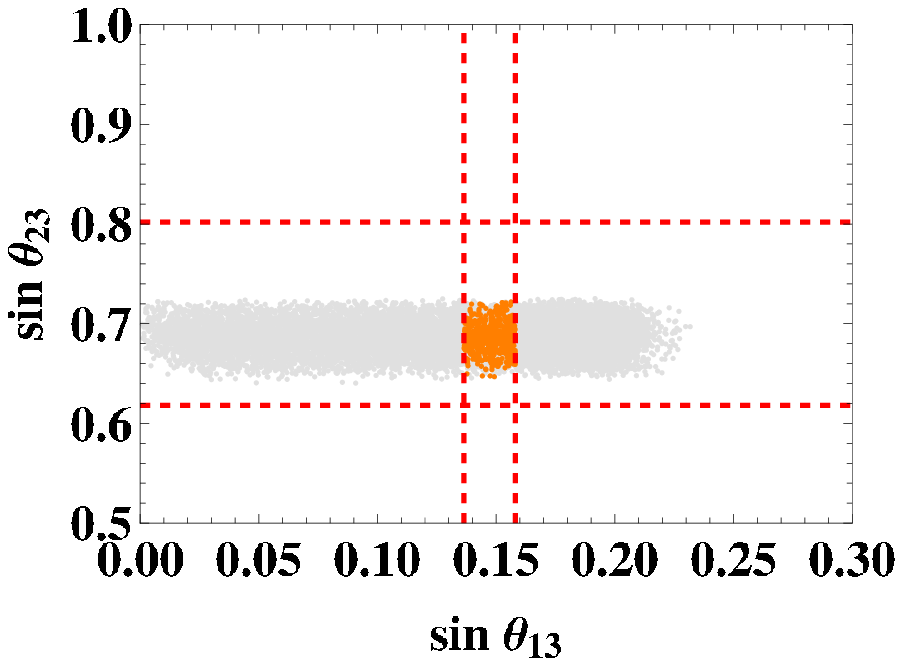}
\end{minipage}
\begin{minipage}{0.32\hsize}
\centering
\includegraphics[clip, width=\hsize]{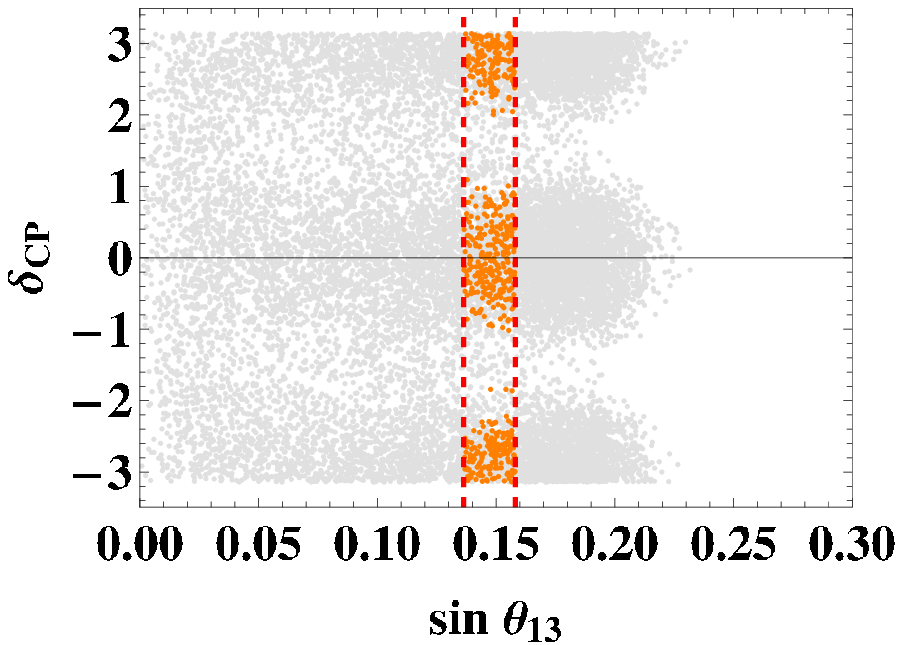}
\end{minipage}
\end{tabular}
\caption{Scatter plots in {Pattern} 4: $a_i=\{\frac{3}{2},\frac{1}{2},0\}$, $c_i=\{0,\frac{1}{2},\frac{3}{2}\}$. We set $\rho =0.5$. 
The color convention is {as in} Fig.~\ref{fig:pattern1}.}
\label{fig:pattern4}
\end{figure}

\begin{figure}[H]
\begin{tabular}{c}
\begin{minipage}{0.32\hsize}
\centering
\includegraphics[clip, width=\hsize]{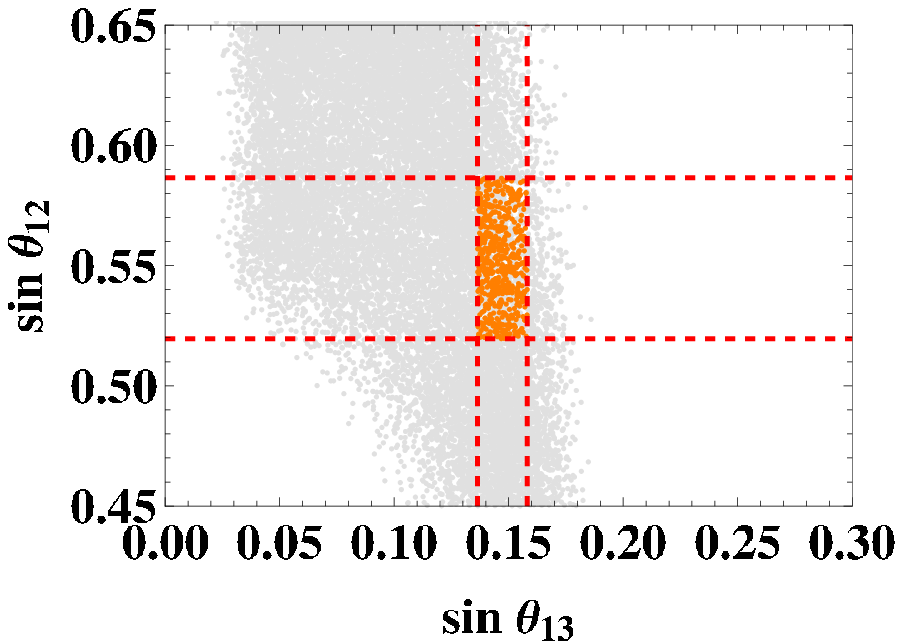}
\end{minipage}
\begin{minipage}{0.32\hsize}
\centering
\includegraphics[clip, width=\hsize]{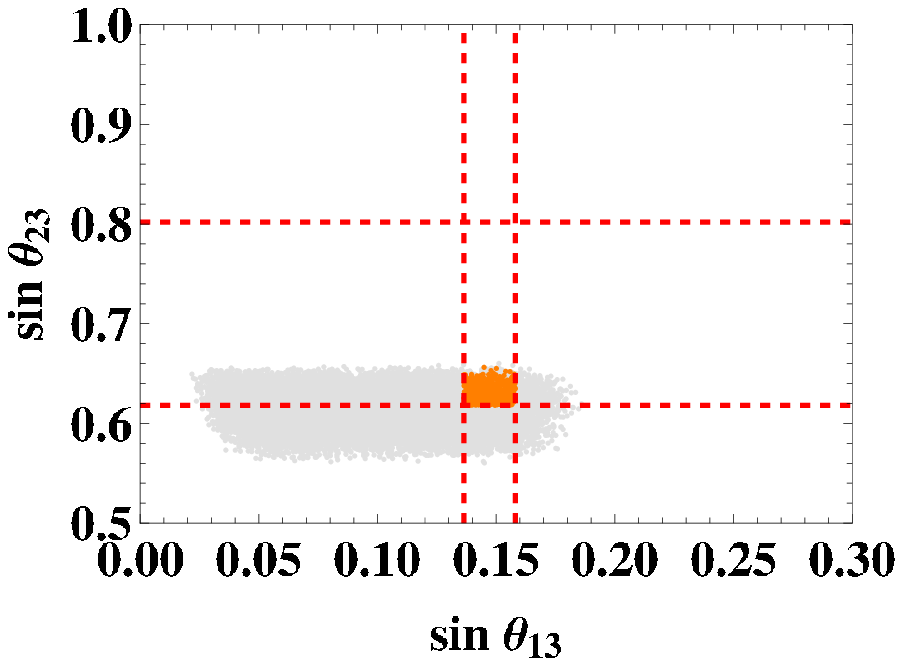}
\end{minipage}
\begin{minipage}{0.32\hsize}
\centering
\includegraphics[clip, width=\hsize]{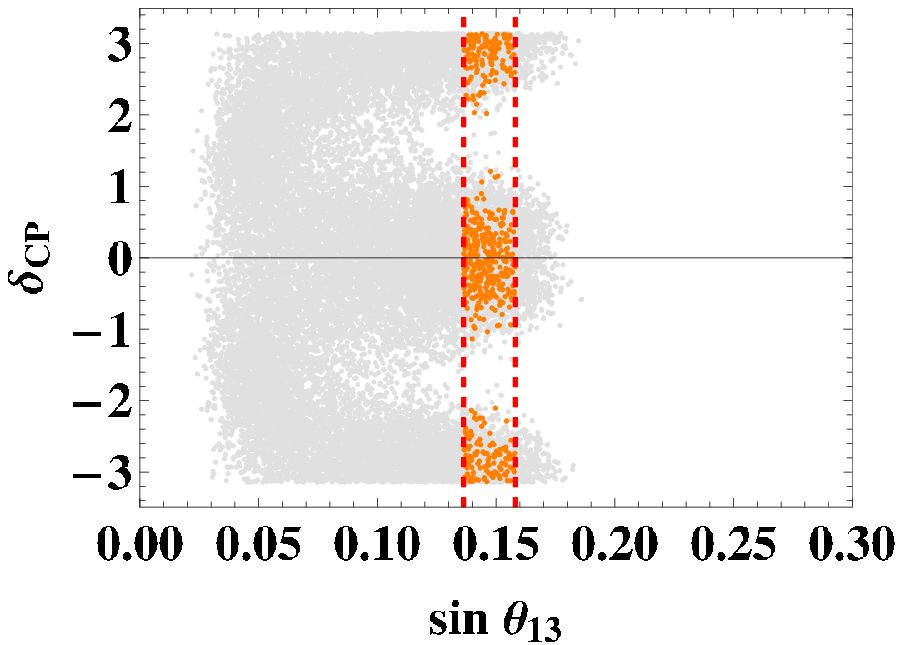}
\end{minipage}
\end{tabular}
\caption{Scatter plots in {Pattern} 5: $a_i=\{\frac{3}{2},\frac{1}{2},0\}$, $c_i=\{0,\frac{1}{2},\frac{3}{2}\}$. We set $\rho =0.6$. 
The color convention is {as in} Fig.~\ref{fig:pattern1}.}
\label{fig:pattern5}
\end{figure}

Next, we show other patterns from {non-$\mu$--$\tau$} symmetric neutrino mass matrices.  
In {Patterns} 3, 4, and 5, the charge configurations of $a_i$ and $c_i$ are $a_i=\{\frac{3}{2},\frac{1}{2},0\}$, $c_i=\{0,\frac{1}{2},\frac{3}{2}\}$, 
while the magnitudes of $\rho $ are {different:} $\rho =0.4,~0.5,~0.6$, respectively. 
If we set $\rho =1.0$ which is the standard FN parametrization, we cannot find the correct ratio of {the} two neutrino mass squared differences and three mixing angles. 
In Figs.~\ref{fig:pattern3},~\ref{fig:pattern4}, and~\ref{fig:pattern5}, the allowed regions of $\sin \theta_{12}$, 
$\sin \theta_{13}$, and {$\delta_{\text{CP}}$} are almost the same, while $\sin \theta_{23}$ is completely different. 
In Figs.~3, 4 and 5, it is easily found that different values of $\rho$ lead to different values of $\sin \theta_{23}$.
This is a remarkable property in the DFN parametrizations.
{Figure}~\ref{fig:pattern3} shows that $\sin \theta _{23}$ is scattered around the upper boundary of the {$3\,\sigma$} range in {Pattern} 3. 
In {Pattern} 4, the allowed region of $\sin \theta _{23}$ is $0.63\lesssim \sin \theta _{23}\lesssim 0.72${,} as shown in Fig.~\ref{fig:pattern4}. 
In {Pattern} 5, Fig.~\ref{fig:pattern5} {shows} that $\sin \theta _{23}$ is around the lower boundary of the {$3\,\sigma$} range. 
When we set $\rho=0.3$ or $0.7$, the obtained values of $\sin \theta_{23}$ are beyond the {$3\,\sigma$} experimental upper and lower bounds, respectively.
The three patterns are tested by measuring the value of $\sin \theta_{23}$ more precisely.
\begin{figure}[H]
\begin{tabular}{c}
\begin{minipage}{0.32\hsize}
\centering
\includegraphics[clip, width=\hsize]{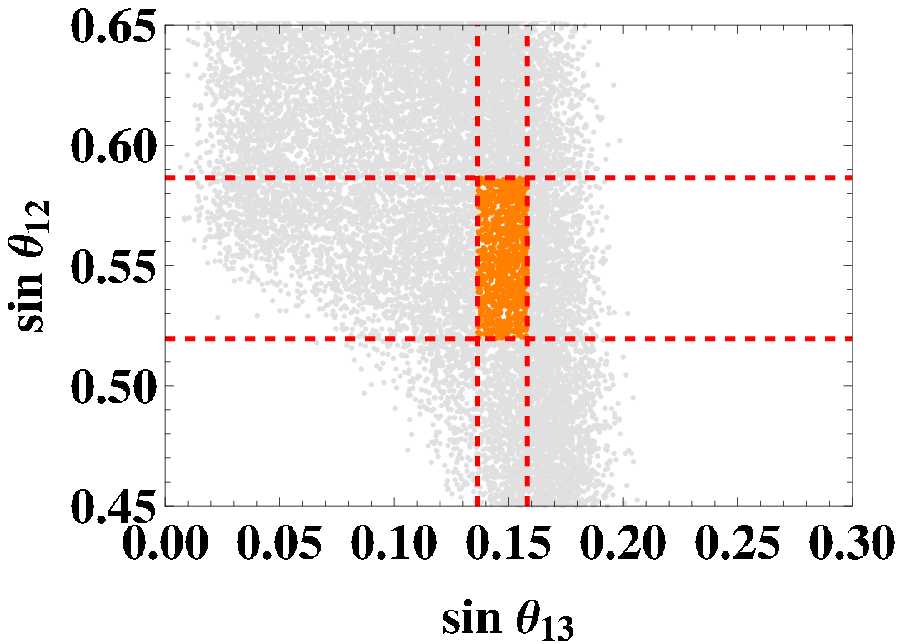}
\end{minipage}
\begin{minipage}{0.32\hsize}
\centering
\includegraphics[clip, width=\hsize]{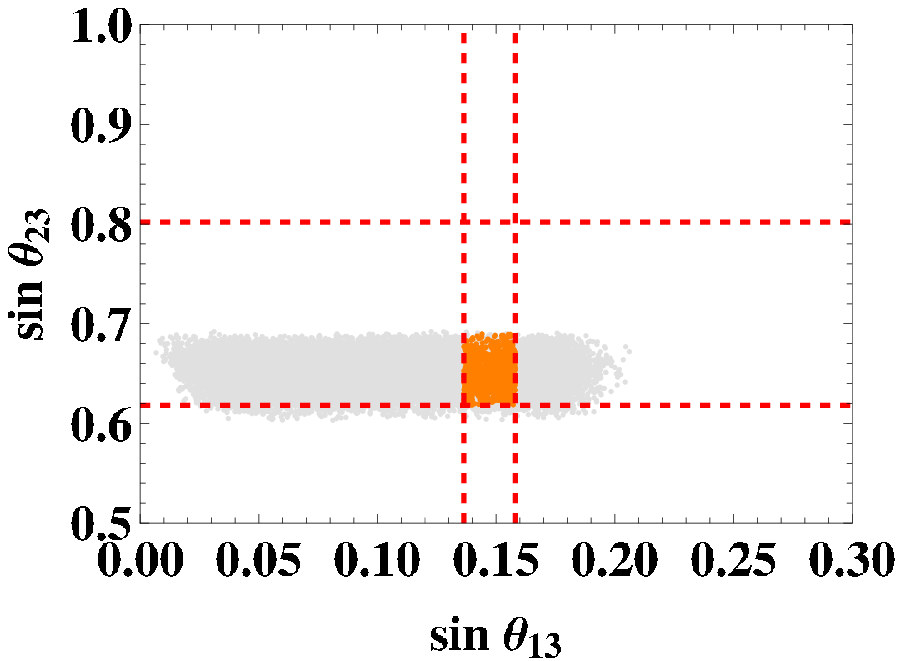}
\end{minipage}
\begin{minipage}{0.32\hsize}
\centering
\includegraphics[clip, width=\hsize]{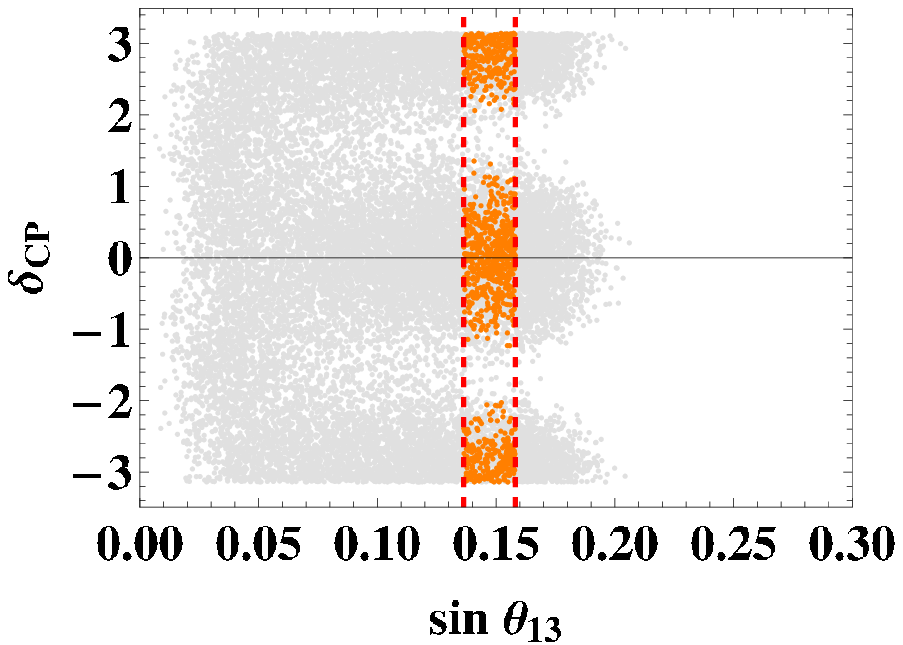}
\end{minipage}
\end{tabular}
\caption{Scatter plots in {Pattern} 6: $a_i=\{\frac{1}{2},\frac{1}{2},0\}$, $c_i=\{1,0,\frac{1}{2}\}$. We set $\rho =0.3$. 
The color convention is {as in} Fig.~\ref{fig:pattern1}.}
\label{fig:pattern6}
\end{figure}

Finally, we show the last pattern. In {Pattern} 6 we set $\rho =0.3${,} which means that this pattern seems to be almost the standard FN parametrization 
because of $\lambda =0.225$.\footnote{If we set $\rho =1.0$, we cannot find the correct neutrino masses and mixing angles 
in the charge configurations of $a_i$ and $c_i$ for {Pattern} 6. Then the DFN parametrization can be well parametrized in the neutrino sector.} 
In Fig.~\ref{fig:pattern6}, $\sin \theta_{12}$ and $\sin \theta_{13}$ are filled within the {$3\,\sigma$} range, 
while $\sin \theta _{23}$ is distributed around the lower boundary of the {$3\,\sigma$} range. 
Note that the neutrino mass matrix in {Pattern} 6 is considered to be similar to that in {Pattern} 1, because the value of {the} additional parameter is small and approximately equal to {$\lambda$,} i.e., $\rho=0.3$.
However, the {values obtained for the} mixing angles are distinct from each other.
Indeed, the DFN extension can make values of $\sin \theta_{13}$ larger and values of $\sin \theta_{23}$ relatively smaller. 
We recognize that these properties are distinctive from those of {Pattern} 1 (the original FN). 
In addition, we find that {$\delta _{\text{CP}}$} is predicted as $\left | {\delta _{\text{CP}}} \right |\lesssim 1$ and 
$2.0\lesssim \left | {\delta _{\text{CP}}} \right |\lesssim \pi$ for {Pattern} 6.

{We comment on the testabilities of the} configurations of the (double) FN charges and parameters.
First, the {values obtained for} $\sin \theta_{23}$ are almost the same in  {Patterns} 1 and 4.
The frequency of consistent values of $\sin \theta_{13}$ is certainly improved in {Pattern} 4.
However, the predicted values are dependent on the coefficients {in front of} each element.
Hence, it is difficult to distinguish {Patterns} 1 and 4 by neutrino experimental data.
Indeed, such situations {happen} between the other patterns, e.g., between {Patterns} 2 and 3 and between {Patterns} 5 and 6.
These coincident properties of predicted mixing angles {can also be seen} in the standard FN parametrizations.
Therefore, all of {the patterns cannot always be} tested by more precise {determination of the} three mixing angles.
This is more conspicuous when ${\cal O}(1)$ coefficients are randomly scattered in {a} wider range, e.g., $c_{ij} \in [0.8,1.2]$.
{The testability of} between different configurations of (double) FN charges and parameters {is} strongly dependent on concrete model {building}.


\section{Discussions and {summary}}
\label{sec:summary}
In the SM, there are many free parameters especially as Yukawa couplings{,} so that 
there are some ambiguities {in realizing} the quark and lepton mass hierarchies and mixing angles. 
{It is therefore} important to study texture analysis or {a} flavor symmetry model in order to elucidate the origin of the flavor structure as {beyond} the SM. 
{As is well known,} mass matrices of the up- and down-type quarks and the charged leptons in the SM 
can be {parametrized well} by the parameter $\lambda$ which is usually taken to be the sine of the Cabibbo angle ($\lambda = {\sin{\theta_{\text{C}}}} \simeq 0.225$). 
In this parametrization, {the} mass hierarchies and mixing angles of the quarks and charged leptons are reproduced. 
However, in the neutrino sector, there {is} still room to realize {the} neutrino mass squared differences $\Delta m_\text{sol}^2$ and $\Delta m_\text{atm}^2$, 
two large mixing angles $\theta _{12}$ and $\theta _{23}$, and non-zero $\theta _{13}$. 
Actually, if the neutrinos are Majorana particles, in the seesaw mechanism, we need {both the} Dirac and Majorana mass terms. 
Even if the Dirac neutrino mass matrix is parametrized by $\lambda$ like the other SM fermions, 
the Majorana masses include free mass parameters in general, and it is plausible that Majorana masses are parametrized by another parameter. 
{Thus}, in this paper {we have presented a} doubly parametric extension of the FN mechanism {with the parameter} $\rho$ in addition to $\lambda $. 
Taking the relevant FN charges for {the} power of $\lambda ~(=0.225)$ and additional FN charges for {the} power of $\rho${,} which we assume {to be} less than one, 
we can reproduce the ratio of the neutrino mass squared differences and lepton mixing angles. Here we assume that the charged lepton mass matrix is diagonal. 

In our calculations, the neutrino masses, {assuming} the normal neutrino mass hierarchy, 
are adjusted by the ratio of the neutrino mass squared differences because the overall mass scale is completely free for our parametrization. 
Here we take ${\cal O}(1)$ coefficients as {$10\%$} deviations from one and {the} complex phases are taken from $-\pi$ to $\pi$. 
Note that if we take the magnitude of ${\cal O}(1)$ coefficients as {$20\%$} deviations from one, 
the allowed region of {$\delta _{\text{CP}}$} is $-\pi \lesssim {\delta _{\text{CP}}} \lesssim \pi$, 
while if we take the magnitude of ${\cal O}(1)$ coefficients as {$5\%$} deviations from one, {$\delta _{\text{CP}}$} is more predictive. 
In this paper, we considered six patterns with/without additional FN charges as sample patterns {for} numerical calculations. 
First, we showed the standard FN and DFN parametrizations which are almost {$\mu$--$\tau$} symmetric mass matrices in {Patterns} 1 and 2, respectively. 
We found that {$\delta _{\text{CP}}$} is predicted as $\left | {\delta _{\text{CP}}} \right |\lesssim 1$ and 
$2.2\lesssim \left | {\delta _{\text{CP}}} \right |\lesssim \pi$. 
In {Pattern} 2, $\sin \theta_{23}$ is around the upper boundary of the {$3\,\sigma$} range, while the other mixing angles are completely filled within the {$3\,\sigma$} range.

Next, we showed other patterns where the neutrino mass matrices are not {$\mu$--$\tau$} symmetric. 
In {Patterns} 3, 4, and 5, the charge configurations of $a_i$ and $c_i$ are $a_i=\{\frac{3}{2},\frac{1}{2},0\}$, $c_i=\{0,\frac{1}{2},\frac{3}{2}\}$, 
while the magnitudes of $\rho $ are different{:} $\rho =0.4,~0.5,~0.6$, respectively. 
If we set $\rho =1.0${,} which corresponds to the standard FN parametrization, we cannot find the correct ratio of the two neutrino mass squared differences and three mixing angles{,}
so the DFN pattern parametrizes the neutrino sector {well}. 
Finally, we find a sizable deviation in {Pattern} 6, where the magnitude {of $\rho\,(=0.3)$ is a little} away from the FN value $\lambda\,(\simeq 0.225)$.

We had seen several examples in the mass matrix form formulated under the concept of DFN.
We recognized that patterns of the mixing angles and the Dirac CP phase can deviate
from {the} predicted ones in the FN texture.
{The deviations look} distinctive when fluctuations in the elements of the mass matrix are
within $10\%$ {of} unity.
As pointed out in the previous section, the explicit differences between the standard FN and DFN parametrizations tend to appear {particularly} in values of $\sin \theta_{23}$.
Hence, measuring $\sin{\theta_{23}}$ precisely is achievable in the near future, {for example} in neutrino oscillation
experiments, and such improved measurements can {determine how well the DFN texture works}.

\vspace{0.3cm}
\noindent
{\bf Acknowledgement{s}}

YS is supported in part by National Research Foundation of Korea (NRF) Research Grant NRF-2015R1A2A1A05001869. 
This work is supported in part by Grants-in-Aid for Scientific Research {[No.~16J05332 (YS) and No.~16J04612 (YT)]} from the Ministry of Education, Culture, Sports, Science and Technology in Japan.


\appendix 

\section{Neutrino mass matrix for each pattern}
\label{sec:mass_matrices}
We show the neutrino mass matrix $m_\nu ^{(i)}$ for each Pattern $i$, which we discuss in {Sect.}~\ref{sec:analyses}; 
\begin{align}
m_\nu ^{(1)}&=
\begin{pmatrix}
\lambda ^2 & \lambda & \lambda \\
\lambda & 1 & 1 \\
\lambda & 1 & 1
\end{pmatrix},\quad 
m_\nu ^{(2)}=
\begin{pmatrix}
\lambda ^2 & \lambda \rho ^\frac{3}{2} & \lambda \rho ^\frac{5}{2} \\
\lambda \rho ^\frac{3}{2} & \rho ^3 & \rho ^4 \\
\lambda \rho ^\frac{5}{2} & \rho ^4 & \rho ^5
\end{pmatrix},\quad 
m_\nu ^{(3,4,5)}=
\begin{pmatrix}
\lambda ^3 & \lambda ^2\rho ^\frac{1}{2} & \lambda ^\frac{3}{2} \rho ^\frac{3}{2} \\
\lambda ^2\rho ^\frac{1}{2} & \lambda \rho & \lambda ^\frac{1}{2}\rho ^2 \\
\lambda ^\frac{3}{2} \rho ^\frac{3}{2} & \lambda ^\frac{1}{2}\rho ^2 & \rho ^3
\end{pmatrix}, \nonumber \\
m_\nu ^{(6)}&=
\begin{pmatrix}
\lambda \rho ^2 & \lambda \rho & \lambda ^\frac{1}{2}\rho ^\frac{3}{2} \\
\lambda \rho & \lambda & \lambda ^\frac{1}{2}\rho ^\frac{1}{2} \\
\lambda ^\frac{1}{2}\rho ^\frac{3}{2} & \lambda ^\frac{1}{2}\rho ^\frac{1}{2} & \rho ^\frac{3}{2}
\end{pmatrix},
\end{align}
where the overall mass parameter is omitted for each mass matrix.



\end{document}